# The Sensitivity of Bayesian Kernel Machine Regression (BKMR) to Data Distribution: A Comprehensive Simulation Analysis

Kazi Tanvir Hasan[1], Gabriel Odom[1], Zoran Bursac[1], Boubakari Ibrahimou[1*]

1. Florida International University, Robert Stempel College of Public Health & Social Work, Department of Biostatistics, 11200 S.W. 8th Street, AHC5, Miami, FL 33199

* Corresponding Author:

Boubakari Ibrahimou, PhD

Associate Professor Florida International University

Robert Stempel College of Public Health & Social Work Department of Biostatistics 11200 S.W. 8th Street, AHC5 – 465, Miami, FL 33199.

Email: bibrahim@fiu.edu

## Abstract

Bayesian Kernel Machine Regression (BKMR) has emerged as a powerful tool to detect negative health effects from exposure to complex multi-pollutant mixtures. However, its performance is degraded when data deviate from normality. In this comprehensive simulation analysis, we show that BKMR's power and test size vary under different distributions and covariance matrix structures. Our results demonstrate specifically that BKMR's robustness is influenced by the response's coefficient of variation (CV), resulting in reduced accuracy to detect true effects when data are skewed. Test sizes become uncontrolled (>0.05) as CV values increase, leading to inflated false detection rates. However, we find that BKMR effectively utilizes off-diagonal covariance information corresponding to predictor interdependencies, increasing statistical power and accuracy. To achieve reliable and accurate results, we advocate for scrutiny of data skewness and covariance before applying BKMR, particularly when used to predict cognitive decline from blood/urine heavy metal concentrations in environmental health contexts.

## Introduction

In recent years, there has been a growing emphasis on analyzing the health effects of exposure to complex multi-pollutant mixtures, driven by heightened concerns about the cumulative impact of pollutants on human health[1,2]. This concern spans diverse scenarios, including air pollution, toxic waste, persistent organic chemicals, and the interplay between environmental exposures and psychosocial factors[1,3,4]. The intricate exposure-response relationships from such mixtures, characterized by their non-linear and non-additive nature, pose substantial challenges for conventional analytical methods[5,6]. The Bayesian Kernel Machine Regression (BKMR) model has emerged as a potent tool for solving and understanding these complex relationships[5–7].

The challenges inherent in studying multi-pollutant mixtures are multifaceted. First, the relationships between mixture components and health outcomes are often highly non-linear and non-additive, demanding a flexible modelling approach capable of accommodating such complexities[5,6,8]. Second, the intricate correlation structures between mixture components further compound the analytical challenges[9]. Existing approaches, including clustering methods, statistical learning algorithms, and variable selection techniques must be revised to address these difficulties[10–12]. In contrast, BKMR offers a novel and promising approach to overcome these limitations, providing more accurate and nuanced insights into exposure-response relationships[5,6,13–15].

The BKMR model has proven effective and adaptable in various environmental health settings when determining complex exposure-response connections. For instance, a population-based cohort study conducted by Li et al. used BKMR to explore the effects of air pollutant mixtures, finding a solid link between pollutant mixtures and all-cause mortality, with fine particulate matter (PM2.5) playing a critical role[16]. Zhang et al. used BKMR to investigate the relationship between a mixture of pollutants and overall obesity, finding significant relationships when all chemicals surpassed 60th sampling percentiles[17]. Liang et al. used BKMR to investigate the relationship between several metals and hearing loss, finding a significant positive correlation between metals and hearing impairment[18]. Surabhi Shah-Kulkarni et al. used BKMR analysis to identify lead exposure during late pregnancy as a significant driver of poor developmental indices at six months[19]. Luli Wu et al. used BKMR to investigate complex relationships between many factors and neural injury biomarkers[20]. These examples highlight the flexibility and power of the BKMR model in unraveling complex exposure-response relationships in various environmental health studies.

Despite its numerous advantages, the BKMR model has limitations. Specifically, BKMR may encounter challenges in producing accurate and stable results when dealing with non-normally distributed data[5,6]. This emphasizes the importance of being mindful of potential challenges when utilizing BKMR and taking steps to address them to ensure the reliability and validity of findings. Another notable drawback is the absence of a specified cutoff point for the Posterior Inclusion Probability (PIP) in determining variable importance, as Bobb et al. highlighted in their original paper[5]. While they mentioned that higher PIP values indicate higher variable importance, they did not provide a threshold for discerning the significance of these probabilities as necessary for hypothesis testing or empirical decision support. Many researchers have used an uninformed PIP threshold of 0.5 to draw inferences regarding variable importance,[15–17,21–24] without considering specific characteristics of their data.

Our research involved a comprehensive simulation analysis to deepen our understanding of the behavior of BKMR and its results. Notably, we found that the estimation of BKMR results is highly sensitive to the synthetic data distribution. Specifically, when data are drawn from a multivariate skewed gamma distribution rather than a multivariate Gaussian distribution, the BKMR results become inaccurate, exhibiting both inflated test size in some situations and decreased power in others. This sensitivity emphasizes the importance of considering data distribution when working with complex models like BKMR to ensure reliable, accurate, and consistent results. Moreover, our findings suggest that relying on a fixed PIP threshold of 0.5 to infer variable importance may lead to overlooking relevant variables associated with the outcome or to including irrelevant variables by mistake. The skewed data distribution introduces complexities that challenge the appropriateness of this conventional threshold, emphasizing the need for a subtle understanding of variable importance determination in skewed data scenarios. Our study will show the possible difficulties associated with using a fixed PIP threshold and that BKMR model performance is not consistent across different data distributions.

## Methodology

### Overview to BKMR:

Bayesian Kernel Machine Regression (BKMR) is a statistical method for modelling the relationship between a response variable and one or more predictor variables without assuming a specific functional form. BKMR is a robust method that accounts for non-linearity and non-additive mixtures by allowing for flexible and complex relationships between exposure variables and health outcomes. The core idea of BKMR lies in using kernel functions, notably the Gaussian kernel, to represent the exposure-response relationships. This choice is motivated by the Gaussian kernel's conjugate prior distribution property, which reduces the computational cost of Bayesian inference. This means that the Gaussian kernel can be easily integrated into a Bayesian framework to estimate model parameters and perform inference on the regression coefficients[5,6,13,14,23][5,6,13,14,22]. The Gaussian kernel represents the relationships between exposure variables and health outcomes flexibly and adaptively, capturing a wide range of potential exposure-response patterns[5,6,14].

### Model Equation:

The BKMR model is an extension of the Kernel Machine Regression model that allows for including a mixture component[5,6]. The BKMR model is given by

$$Y_i = h(z_i1, \ldots, z_iM) + x_i'\beta + \gamma_i + \epsilon_i,$$

Depending on the desired approach, the BKMR model can be fit using maximum likelihood estimation or Bayesian methods[5,6].

### Model Components:

The BKMR model is built on several key components[5,6]-

- Health Outcome (Response Variable): Denoted as $Y_i$, this represents the health measurement for the i-th individual in the study.

- Exposure Variables (Predictors): Represented by the vector $z_i = (z_i1, z_i2, \ldots, z_iM)$, where M is the number of exposure variables. These could be pollutants, chemicals, or other environmental factors known to impact health.

- Exposure-Response Function: Represented by $h(z)$, this function characterizes how the health outcome changes in response to different levels of exposure. BKMR assumes that this function is smooth and can capture complex nonlinearities.

- Kernel Function: The kernel function, often the Gaussian kernel, captures the similarity or distance between different exposure profiles. It plays a crucial role in estimating the exposure-response function.

- Confounding Variables: Represented by the vector $x_i$, these variables help control for potential confounding factors that might affect the relationship between exposures and the health outcome, such as demographics, socio-economic status, co-morbid diseases conditions, etc.

- Regression Coefficients: Denoted by $\beta$, these coefficients determine the strength and direction of the linear relationship between the confounding variables and the health outcome.

- Error Term: Denoted by $\epsilon_i$, this captures the variability or noise in the relationship that the model cannot explain.

## Simulation Analysis:

To preserve the correlation patterns observed in real-world datasets, we generated simulated data using a real dataset as a pattern. This section details the steps involved in generating simulated data for testing the performance of the BKMR model in various scenarios.

1. Data Acquisition and Preparation:
- Source: National Health and Nutrition Examination Survey (NHANES) data spanning 2011-2014[25,26].

- Data types: Separate datasets were acquired for demographics, medical conditions, cognitive function assessments, and blood metal concentrations.

- Merging: Individual datasets were merged into a single comprehensive dataset for further analysis.

2. Data Transformation:
- Target variables: Metal concentration variables (lead, cadmium, mercury, selenium, manganese) underwent logarithmic transformation (log10(x+1)) to reduce skewness.

- Standardization: Transformed metal variables were further scaled using their respective standard deviations to bring them on a common scale and facilitate meaningful comparisons.

3. Sex-Specific Segmentation:
- To account for potential sex differences in covariance matrix structures, the dataset was segmented into separate male and female subsets.

- This segmentation allows for more accurate synthetic data generation, as we observed different correlation pattern between exposures and confounders by sex.

4. Parameter Estimation:
- For each sex-specific data subset, the following parameters were estimated for the transformed metal variables:

    - Mean (μ)

    - Standard deviation (σ)

    - Covariance matrices ($\Sigma\_d, \Sigma\_u$), a diagonal covariance matrix and an unstructured covariance matrix, respectively. The real data features have correlation values ranging from 0.00 to 0.34, so simulating data using a diagonal covariance matrix allows us to inspect the BKMR method performance when no feature mixtures exist.

    - Gamma distribution parameters (α, β) using the dampack::gamma_params() function to capture the underlying skewed distribution characteristics[27].

- New gammaParam() function was designed to efficiently calculate parameters while excluding missing values within each sex group.

- We estimated the metal distribution parameters using a subset of the original data that had complete cases for blood metals (2934 sample total, 2068 with complete blood metals data).

5. Parametric Bootstrap Simulation of Metal Data:
- Utilizing the estimated parameters from step 4, we simulated 100 independent samples of size N = 2934 from the Multivariate Skewed Gamma distribution (using lcmix:: rmvgamma) for all metal variables for both males and females, with both unstructured and diagonal covariance matrices. We chose the Multivariate Skewed Gamma distribution because it provides a more realistic portrayal of the underlying sufficient statistics and correlations between metal variables while adjusting for the skewness commonly found in environmental exposure data[28].

- The number of simulated samples per sex (1506 for females and 1428 for males) was chosen to match the proportions in the original NHANES dataset, ensuring a representative sample size.

- The selection of this distribution corresponds to the properties of the measured metal concentrations and ensures that the simulated samples retain important characteristics of the original data, such as mean, standard deviation, skewness, and covariance matrices. This step in the simulation process is intended to preserve the complexity of the real data structure and contribute to a more accurate evaluation of the BKMR method's performance under realistic settings.

- These simulated metal concentration values for both sexes were then combined to form 100 final simulated data sets with 2934 observations, mirroring the size of the original NHANES data.

6. Sample Generation and Comparison:
- To assess the accuracy and reliability of the simulated data, we compared the characteristics of the 100 simulated samples for each metal to those of the original NHANES data.

- The comparison metrics included: Mean, Standard Deviation, Skewness, and Equality of Covariance matrices.

- This detailed comparison ensured that the simulated data preserved, as much as possible, the essential statistical properties of the real-world data, including the complex interrelationships between the metal variables.

7. Effect Size Simulation:
- Bobb et al. in their original paper simulation only generated multivariate normal response data with a fixed variance parameter and a single correlation matrix structure[5]. In contrast, our simulation design employs a multivariate skewed gamma distribution, a range of 12 variance values (with coefficient of variation values ranging from 0.02 to 15 in absolute value), and also we compare the effects of diagonal and unstructured covariance matrices. All of these additions fill gaps in the literature.

- To measure the type-I error rate of BKMR (the probability of rejecting the null hypothesis when the null hypothesis is true-i.e., the simulated data should show no relationship between the metals and cognitive function), we simulated using the following two methods for both unstructured covariance and diagonal covariance data:

Method 1: Varying Means with Normal Distribution -

- We used a normal distribution to generate response data with different mean values.

- Parameters: Mean values were set as follows: -4.25, -2.4, -1.9, -1.61, -1.27, -0.70, -0.50, -0.30, -0.223, -0.19, -0.08, -0.05, -0.025, -0.01, 0.94, all with a fixed standard deviation of 0.10. These values were chosen to span the range of realistic values based on previous literature[29-33].

Method 2: Varying Standard Deviations with Normal Distribution

- Response data was generated using a normal distribution with a fixed mean and varying standard deviations.

- Parameters: The mean was set at -1.0, while standard deviations were set as follows: 0.10, 0.25, 0.50, 0.67, 1.0, 1.5, 2.0, 3.5, 5.0, 7.5, 10.0, and 15.0. These values were chosen to adequately test the performance of the BKMR in realistic and extreme scenarios.

- The simulated metal concentration data features have various skewness values ranging from 0.006296 to 1.534478 (original data skewness values are ranging from 0.4 to 1.55), allowing us to measure the effect of departures from multivariate normality on the BKMR method. Bobb et al. in their original paper simulation only generated from multivariate normal data, so this simulation design, which employs data from a multivariate skewed gamma distribution, fills a gap in the literature[5].

- This simulation design focuses exclusively on untreated data, allowing us to investigate the probability of the BKMR method to erroneously reject the null hypothesis due to random chance. This design allowed us to evaluate the ability of BKMR to recover the true effect sizes of the simulated metal exposures on various outcome variables. To the best of our understanding, the original papers did not perform a simulation study to assess the test size of the BKMR method[5].

8. Power Analysis Simulation:
- Random Seed and Noise Generation Random Seed Initialization:
    - A random seed was established to ensure reproducibility.
    - Error Component: Random noise ($\epsilon$) was generated from a normal distribution: $\epsilon \sim N(\mu=0, \sigma=2)$.

- Signal Definition: In this study "signals" were defined based on blood lead and mercury concentrations. In real-world data after the transformation described above (log transformation followed by scaling), the distribution of lead can be approximated by a normal distribution with a slightly heavier tail on the right. However, the distribution of mercury is highly skewed to the right, even after the transformation above. We chose mercury instead of cadmium because although cadmium also exhibits right-skewness, existing literature suggests that mercury has a stronger association with cognitive decline compared to cadmium.

Normal Signals: Three levels of normal signal were created:

- NormalLow = - 0.03* Lead
- NormalMedium = - 0.08 * Lead
- NormalHigh= - 0.12 * Lead

Skewed Signals: Three levels of skewed signal were generated:

- SkewedLow = - 0.02 * Mercury
- SkewedMedium = - 0.10 * Mercury
- SkewedHigh = - 0.15 * Mercury

Interaction Only Signals: Three interaction signal levels were introduced:

- InteractionLow = - 0.01 * Lead * Mercury
- InteractionMedium = - 0.05 * Lead * Mercury
- InteractionHigh = - 0.10 * Lead * Mercury

Full Factorial Signals: Three levels were generated based on the combination of lead, mercury and their interactions:

- FullFactorialLow = 0.5 * (- 0.03 * Lead - 0.04 * Mercury - 0.01 * Lead * Mercury)
- FullFactorialMedium = 0.5 * (- 0.07 * Lead - 0.06 * Mercury - 0.03 * Lead * Mercury)
- fullFactorialHigh = 0.5 * (- 0.12 * Lead - 0.08 * Mercury - 0.05 * Lead * Mercury)

Signal Modification: The previously generated random noise was added to each signal category (normal, skewed, interaction, and full factorial)

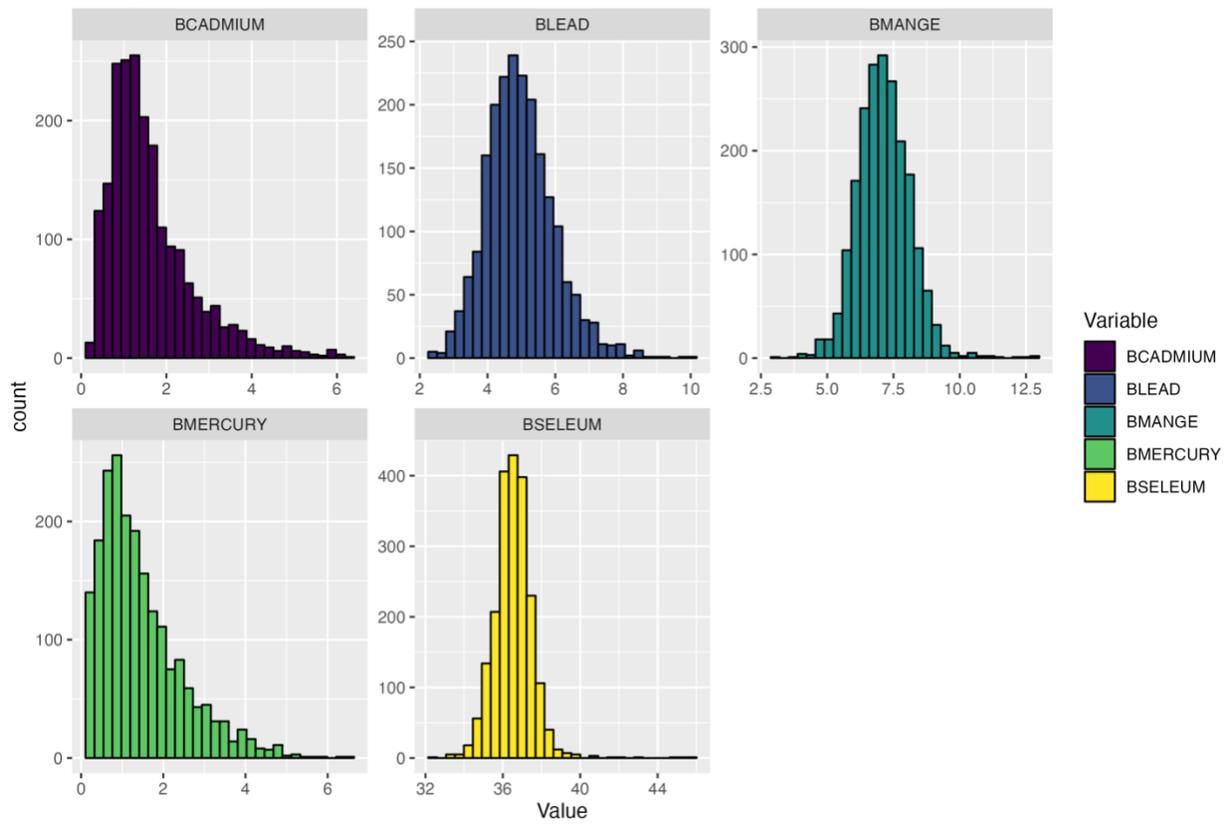

9. BKMR Model Fitting and Evaluation:
- 100 simulated datasets (for both true unstructured covariance matrix and diagonal covariance matrix) were used to evaluate the performance of BKMR under different scenarios.

- The 'kmbayes' function from the R BKMR package was used to fit the BKMR model to each selected dataset using a Markov chain Monte Carlo (MCMC) algorithm with 10000 iterations and component-wise variable selection, as recommended in the original paper[34].

- This rigorous selection and analysis workflow allowed for a comprehensive assessment of BKMR's effectiveness in identifying the true associations between simulated metal exposures and outcomes across various signal strengths.

- In this analysis, we used a high-performance computing (HPC) node. The node is equipped with four Intel(R) Xeon(R) CPU E7-8890 v3 processors, each operating at 2.50GHz, providing a total of 72 CPU cores. Additionally, the node has 512 GB of memory.

# Results

## Effect Size Analysis:

To analyze effect size in this study, we examined how changes in the Coefficient of Variation (CV) affected the results of the BKMR analysis when using diagonal and unstructured covariance matrix data. We made two distinct adjustments to induce changes in the coefficient of variation (CV). Firstly, we fixed the SD at a constant value of 0.1 and adjusted the mean from -4.25 to 0.94. Secondly, we maintained the mean at -1.0 and systematically varied the standard deviation (SD) from 0.10 to 15. These points are chosen to emulate and expand upon design values previously published[29-33].

In the diagonal covariance matrix data, modifying CV values through mean or SD resulted in similar changes in the test size across variables (cadmium, lead, manganese, mercury, and selenium). Smaller CV values resulted in decreased test sizes (less than 0.05), while larger CV values resulted in uncontrolled test sizes (greater than 0.05), indicating the sensitivity of test size to CV differences. Specifically, test sizes were constantly uncontrolled for CV values greater than 2.

In unstructured covariance matrix data, changes in CV also impacted test size, highlighting the sensitivity of effect size to CV differences. However, for this data, test sizes were uncontrolled (greater than 0.05) for CV values greater than 5, implying that the BKMR method employs information in the full unstructured covariance matrix to control the test size. These findings emphasize the importance of considering feature engineering procedures which are appropriate for the chosen statistical method; for example, if the response variable is z-score normalized as a preprocessing step, then the CV is undefined, and the type-I error rate for the BKMR method could exceed 50%.

These results provide valuable insights into BKMR model performance across different data structures and help us better understand the strengths and limitations of the current BKMR formulation and its analysis results.

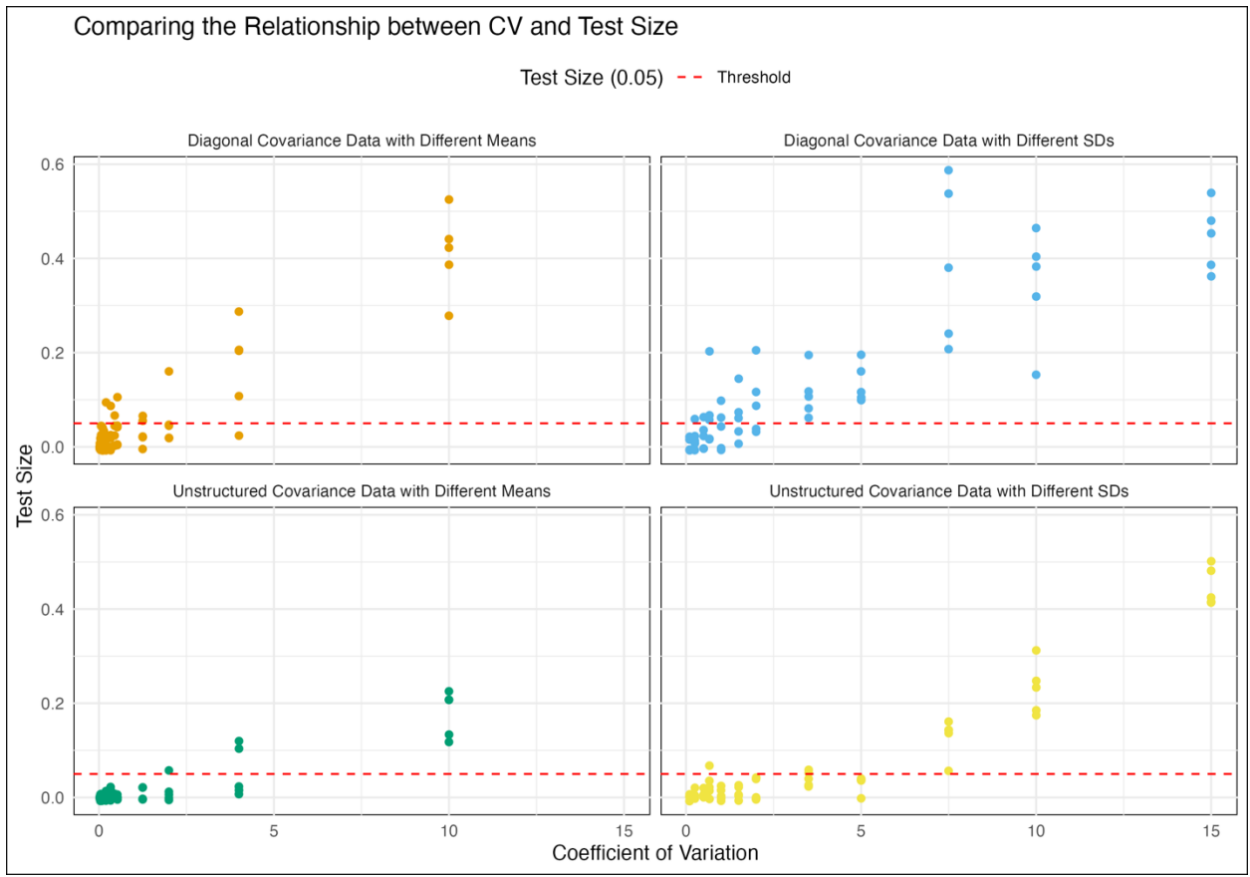

## Unstructured Covariance Matrix:

**Power Analysis:**

We conducted a power analysis to assess the ability of the BKMR model to detect significant effects under different experimental conditions. Varying levels of lead (less skewed, skew = 0.618) and mercury (more skewed, skew = 1.28) characterized the experimental designs. Table 1 presents the statistical power associated with the experimental conditions.

Table 1 contains power analysis results: rows represent a specific experimental design scenario, while the columns correspond to different metals of interest (Cadmium, Lead, Manganese, Mercury, and Selenium). The values in the table represent the statistical power (probability of rejecting the null hypothesis) associated with the experimental conditions. Highlighted statical power values represent the variables for which the null hypothesis should be rejected, that is, these values measure the probability of correctly detecting a significant effect for the respective metal. For example, in Table 1 the Normal High and Skewed High scenarios reveal significant differences in the statistical power and false detection rates of the BKMR model across various metal concentrations. Under the Normal High scenario, the model demonstrates robust power in detecting the treatment effect of lead (power = 0.9). Moreover, the false detection rates for untreated metals, including cadmium, manganese, mercury, and selenium, are relatively low (ranging from 0.0 to 0.1). In contrast, under the Skewed High scenario, while the model retains strong power to detect the true signal from the treated mercury variable (power = 0.9), it exhibits inflated false detection rates for untreated metals. False detection rates are notably higher for cadmium, lead, manganese, and selenium (ranging from 0.3 to 0.4), indicating a higher likelihood of falsely identifying treatment effects. The presence of skewed data prohibits the BKMR method from accurately distinguishing true treatment effects from random fluctuations, which is particularly noticeable in the elevated false detection rates for untreated metals in the Skewed High scenario. Therefore, because the test size is not well controlled, it is inappropriate to use statistical power to evaluate BKMR model performance. Instead, we will discuss classification accuracy.

Table 1: Power Analysis (Unstructured Covariance Matrix) Results for Different Experimental Design

| Design | Cadmium | Lead | Manganese | Mercury | Selenium |
|---|---|---|---|---|---|
| Normal Low | 0.2 | **0.2** | 0.3 | 0 | 0.2 |
| Normal Medium | 0.1 | **0.7** | 0.1 | 0 | 0.1 |
| Normal High | 0 | **0.9** | 0.1 | 0 | 0 |
| Skewed Low | 0.4 | 0.4 | 0.4 | **0.4** | 0.3 |
| Skewed Medium | 0.4 | 0.6 | 0.5 | **0.8** | 0.3 |
| Skewed High | 0.3 | 0.3 | 0.3 | **0.9** | 0.4 |
| Interaction Low | 0.3 | **0.4** | 0.4 | **0.2** | 0.3 |
| Interaction Medium | 0.2 | **0.6** | 0.2 | **1** | 0.3 |
| Interaction High | 0 | **1** | 0.1 | **1** | 0 |
| Full Factorial Low | 0 | **0.1** | 0.1 | **0.2** | 0.1 |
| Full Factorial Medium | 0.2 | **0.6** | 0.3 | **0.7** | 0.3 |
| Full Factorial High | 0 | **0.8** | 0.1 | **1** | 0.3 |

* Bold values are for treated variables.

**Confusion Matrix:**

In contrast to the statistical power in Table 1, the confusion matrix in Table 2 presents a summary of model performance for these experimental designs uninfluenced by BKMR's inflated test size. The matrix displays the proportion of true positives (BKMR correctly detects a treated metal feature), false positives (BKMR incorrectly detects an untreated metal feature), false negatives (BKMR incorrectly fails to detect a treated metal feature), and true negatives (BKMR correctly fails to detect an untreated metal feature). For example, in the "Normal High" scenario, the model correctly identifies lead's effect on simulated cognition in 90% of simulated datasets (which are true positives) while incorrectly identifying effects from untreated metals in only 2% of simulated datasets (false positives). However, this model fails to identify lead's effect on simulated cognition in 10% of simulated datasets (false negatives), but BKMR correctly identifies 98% of simulations with untreated metals as true negatives. For comparison, in the "Skewed High" scenario, the model maintains the same accuracy in identifying actual positive cases. However, it shows a dramatically higher false positive rate (33% vs 2%) and lower true negative rate (68% vs 98%).

Table 2: Confusion Matrix (Unstructured Covariance Matrix) for Each Design Point

| Design | True Positive | False Positive | False Negative | True Negative |
|---|---|---|---|---|
| Normal Low | 0.2 | 0.17 | 0.8 | 0.82 |
| Normal Medium | 0.7 | 0.12 | 0.3 | 0.88 |
| Normal High | 0.9 | 0.02 | 0.1 | 0.98 |
| Skewed Low | 0.21 | 0.19 | 0.79 | 0.81 |
| Skewed Medium | 0.8 | 0.45 | 0.2 | 0.55 |
| Skewed High | 0.9 | 0.33 | 0.1 | 0.68 |
| Interaction Low | 0.38 | 0.58 | 0.62 | 0.42 |
| Interaction Medium | 0.8 | 0.24 | 0.2 | 0.76 |
| Interaction High | 0.95 | 0 | 0.05 | 1 |
| Full Factorial Low | 0.6 | 0.49 | 0.4 | 0.51 |
| Full Factorial Medium | 0.62 | 0.37 | 0.38 | 0.63 |
| Full Factorial High | 0.82 | 0.11 | 0.18 | 0.89 |

In order to efficiently compare model performance across these 12 designs, we summarize the confusion matrices with classification performance statistics: Accuracy $\left(\frac{TP+TN}{TP+TN+FP+FN}\right)$, Precision $\left(\frac{TP}{TP+FP}\right)$, Recall $\left(\frac{TP}{TP+FN}\right)$, F1 Score $\left(2 \times \frac{Accuracy \times Precision}{Accuracy + Precision}\right)$. Table 3 displays evaluation metrics summarizing the model's performance for each experimental design scenario. In the "Normal High" scenario, the model achieves high accuracy (94% correctly classified metals) and near perfect precision (98% of metals detected by BKMR had truly been treated). However, recall, the probability of BKMR detecting a truly treated metal, was lower (90%), indicating that BKMR missed several positive cases. Consequently, the F1 score (the harmonic mean of precision and recall) is reduced (94%). For comparison, in the "Skewed High" scenario, although the recall remains unchanged (90%), the precision is much lower compared to the "Normal High" scenario (73% vs 98%). Therefore, the accuracy (79% vs 94%) and F1 score (81% vs 94%) are notably lower. We summarize these conflicting results as 1) if the BKMR indicates that a metal had been treated, this prediction is less likely to be true if the underlying distribution is skewed; yet 2) the BKMR method can detect truly treated metals with the same power regardless of the skewness of the distribution. Therefore, the BKMR method suffers reduced accuracy when the data cannot be approximated by a multivariate Gaussian distribution.

Table 3: Evaluation Matrix for Each Design Point

| Design | Accuracy | Precision | Recall | F1 Score |
|---|---|---|---|---|
| Normal Low | 0.51 | 0.54 | 0.20 | 0.29 |
| Normal Medium | 0.79 | 0.85 | 0.70 | 0.77 |
| Normal High | 0.94 | 0.98 | 0.90 | 0.94 |
| Skewed Low | 0.51 | 0.53 | 0.21 | 0.30 |
| Skewed Medium | 0.68 | 0.64 | 0.80 | 0.71 |
| Skewed High | 0.79 | 0.73 | 0.90 | 0.81 |
| Interaction Low | 0.40 | 0.40 | 0.38 | 0.39 |
| Interaction Medium | 0.78 | 0.77 | 0.80 | 0.78 |
| Interaction High | 0.98 | 1.00 | 0.95 | 0.97 |
| Full Factorial Low | 0.56 | 0.55 | 0.60 | 0.57 |
| Full Factorial Medium | 0.63 | 0.63 | 0.62 | 0.62 |
| Full Factorial High | 0.86 | 0.88 | 0.82 | 0.85 |

Comparing the recall and true positive values between the experimental designs reveals compelling insights into the model's efficacy in identifying specific metal contaminants. For example, in the Normal High design, the model's ability to consistently identify 90% of the actual positive cases for each metal contaminant (as indicated by the recall value of 0.90) is a significant finding. This is further reinforced by the true positive values, which mirror the recall scores at 0.9 for both lead and mercury. These results underscore the model's balanced performance in accurately detecting lead in the Normal High design and mercury in the Skewed High design, affirming BKMR's crucial role in detecting metal contamination in different scenarios.

However, while the recall metric provides valuable insights into the model's ability to capture positive cases within specific contaminant classes, a noteworthy classification disparity emerges when comparing accuracy and recall within the two designs. In the Normal High design, the model achieves a commendable accuracy of 0.94. However, the recall for lead stands at 0.90, suggesting a potential limitation. This implies that while the model demonstrates a high overall predictive capability, it may overlook a proportion of actual positive cases of lead contamination, capturing only 90% of such instances. Conversely, within the Skewed High (treating mercury) design, the model has a comparatively lower overall predictive accuracy of 0.79. However, the recall for mercury remains constant at 0.90, indicating the model's consistent ability to identify 90% of the actual positive cases of mercury contamination. This highlights a crucial distinction between accuracy and recall: accuracy provides a holistic assessment across all classes, while recall focuses specifically on positive cases within individual classes, disregarding false negatives. In summary, despite differences in overall accuracy between the two designs, the comparable recall values suggest that the model is consistent in its capability to detect treated metals under various data distributions. However, the accuracy suffers significantly, underscoring the importance of considering accuracy and recall metrics simultaneously for comprehensive model evaluation in environmental contamination studies. This performance disparity highlights the need for further research to improve BKMR performance in non-Gaussian scenarios.

## Diagonal Covariance Matrix:

### Power Analysis:

Unlike the unstructured covariance matrix data power analysis (as shown in Table 1), the power analysis employing a diagonal covariance matrix indicates moderate statistical power for the treated metals, using same effect sizes. However, the power to detect untreated metals (false positives) are much higher, indicating that the BKMR method draws considerable information from the covariance matrix to make more accurate decisions.

Table 4: Power Analysis (Diagonal Covariance Matrix) Results for Different Experimental Design

| Design | Cadmium | Lead | Manganese | Mercury | Selenium |
|---|---|---|---|---|---|
| Normal Low | 0.5 | **0.52** | 0.39 | 0.47 | 0.43 |
| Normal Medium | 0.54 | **0.5** | 0.41 | 0.5 | 0.4 |
| Normal High | 0.55 | **0.61** | 0.37 | 0.51 | 0.41 |
| Skewed Low | 0.53 | 0.53 | 0.48 | **0.55** | 0.49 |
| Skewed Medium | 0.59 | 0.56 | 0.5 | **0.62** | 0.51 |
| Skewed High | 0.5 | 0.51 | 0.35 | **0.59** | 0.4 |
| Interaction Low | 0.56 | **0.57** | 0.45 | **0.53** | 0.46 |
| Interaction Medium | 0.65 | **0.66** | 0.61 | **0.64** | 0.57 |
| Interaction High | 0.55 | **0.56** | `0.47 | **0.59** | 0.47 |
| Full Factorial Low | 0.46 | **0.5** | 0.4 | **0.44** | 0.39 |
| Full Factorial Medium | 0.63 | **0.66** | 0.53 | **0.62** | 0.56 |
| Full Factorial High | 0.43 | **0.46** | 0.32 | **0.43** | 0.35 |

* Bold values are for treated variables.

### Confusion Matrix:

Similar to the confusion matrix for unstructured data in Table 2, the confusion matrix in Table 5 presents a summary of model performance on data using a diagonal covariance matrix for these experimental designs uninfluenced by the BKMR inflated test size. Note that these values are all considerably lower than in the unstructured covariance matrix scenario. This further supports the idea that the BKMR method draws information heavily from the covariance matrix to accurately detect treated metals.

Table 5: Confusion Matrix (Diagonal Covariance Matrix) for Each Design Point

| Design | True Positive | False Positive | False Negative | True Negative |
|---|---|---|---|---|
| Normal Low | 0.52 | 0.45 | 0.48 | 0.55 |
| Normal Medium | 0.50 | 0.46 | 0.50 | 0.54 |
| Normal High | 0.61 | 0.46 | 0.39 | 0.54 |
| Skewed Low | 0.55 | 0.51 | 0.45 | 0.49 |
| Skewed Medium | 0.62 | 0.54 | 0.38 | 0.46 |
| Skewed High | 0.59 | 0.44 | 0.41 | 0.56 |
| Interaction Low | 0.55 | 0.49 | 0.45 | 0.51 |
| Interaction Medium | 0.65 | 0.61 | 0.35 | 0.39 |
| Interaction High | 0.58 | 0.50 | 0.43 | 0.50 |
| Full Factorial Low | 0.47 | 0.42 | 0.53 | 0.58 |
| Full Factorial Medium | 0.64 | 0.57 | 0.36 | 0.43 |
| Full Factorial High | 0.45 | 0.37 | 0.56 | 0.63 |

Similar to Table 5, the results presented in Table 6 indicate variations in performance across different design points. These findings underscore the importance of considering the dataset's characteristics and the design parameters when optimizing classification performance.

Table 6: Evaluation Matrix for Each Design Point

| Design | Accuracy | Precision | Recall | F1 Score |
|---|---|---|---|---|
| Normal Low | 0.54 | 0.54 | 0.52 | 0.53 |
| Normal Medium | 0.52 | 0.52 | 0.50 | 0.51 |
| Normal High | 0.58 | 0.57 | 0.61 | 0.59 |
| Skewed Low | 0.52 | 0.52 | 0.55 | 0.53 |
| Skewed Medium | 0.54 | 0.53 | 0.62 | 0.57 |
| Skewed High | 0.58 | 0.57 | 0.59 | 0.58 |
| Interaction Low | 0.53 | 0.53 | 0.55 | 0.54 |
| Interaction Medium | 0.52 | 0.52 | 0.65 | 0.58 |

| Interaction High | 0.54 | 0.54 | 0.58 | 0.56 |
| Full Factorial Low | 0.53 | 0.53 | 0.47 | 0.50 |
| Full Factorial Medium | 0.53 | 0.53 | 0.64 | 0.58 |
| Full Factorial High | 0.54 | 0.55 | 0.45 | 0.49 |

The BKMR method in the unstructured matrix scenario has access to additional information concerning the relationships between metals, whereas in this diagonal matrix scenario the data generating process removes potential relationships from the metals data (a limiting assumption which implies that metal exposures are independent of each other). We believe that this loss of information is the reason that the BKMR model accuracy has dropped significantly when ignoring intra metal correlation.

## Discussion

The comprehensive simulation analysis conducted in this study sheds light on the sensitivity of Bayesian Kernel Machine Regression (BKMR) to a data distribution's departure from multivariate normality, particularly when dealing with complex, skewed multi-pollutant mixtures. We believe that the commonly prescribed 0.5 threshold of significance is inappropriate for analysis of skewed data. Our findings underscore the importance of considering the underlying data distribution in BKMR analysis to ensure reliable and accurate results.

The effect size analysis revealed a notable sensitivity of the BKMR method to changes in the Coefficient of Variation (CV) under data simulated from diagonal or unstructured covariance matrices. We believe that this sensitivity indicates that caution should be used when interpreting BKMR analysis outcomes if 1) the underlying data are not multivariate Gaussian and/or 2) correlations and interactions between the predictors are weak or not well-known a priori. Moreover, the power analysis elucidated how varying experimental conditions and data distributions influence the model's ability to detect treatment effects. The results emphasize the importance of understanding the interplay between data distribution, expected predictor interactions, and BKMR performance.

Our findings indicate that BKMR exhibits sufficient power to detect significant effects of treated metals (e.g., lead and mercury) under medium to high signal scenarios. However, the model's performance deteriorates as the distribution of the simulated data departs from normality. Notably, skewed data distributions can lead to inflated false detection rates for untreated metals, highlighting the need to consider data distribution in BKMR analysis carefully. The confusion matrix and evaluation metrics further elucidate the model's performance under different experimental conditions. These metrics provide valuable insights into BKMR's accuracy, precision, recall, and F1 score across various scenarios.

The observed deviations in our analysis may be due to various factors, among which the skewness of the variables holds significance. Our initial examination indicates that metals such as lead, manganese, and selenium exhibit moderate levels of skewness, while cadmium and mercury display noticeable skewness, even after transformation and scaling. This difference in skewness across variables could significantly impact the accuracy of analysis. Furthermore, when considering the design matrix, higher moments beyond mean and variance, such as skewness, can further complicate the modelling process, reducing accuracy. The choice between a diagonal and unstructured covariance matrix for predictor data is also essential. While a diagonal matrix is easier to estimate, it erroneously assumes independence among predictors. However, an unstructured matrix has the flexibility to capture important correlations. Loss of off-diagonal elements in the covariance matrix can undermine the model's ability to account for interdependencies among predictors, thereby impacting accuracy and power. Moreover, the location and scale of the response variable play a crucial role in model performance. As the coefficient of variation (CV) increases, indicating more significant variability relative to the mean, accuracy tends to suffer. Therefore, it is essential to consider the distributional properties of variables, covariance structure, and response variable characteristics to ensure the BKMR results are robust and accurate.

Because of our findings, we believe that a fruitful area of future research would include: a mathematical inspection of the BKMR method to investigate its sensitivity to the coefficient of variation of a response vector; a thorough simulation study to map out the relationship between summary statistics and moments from the data and threshold values which preserve a nominal type-I error rate of 5% for BKMR; and potential reanalysis of previously published BKMR results using updated and appropriate threshold values. As we mentioned previously, many researchers have commonly used a PIP threshold of >0.5 to draw inferences regarding variable importance, a practice we found problematic in our simulation studies, especially in skewed data scenarios and scenarios with multiplicative interaction effects, which we will inspect in future research.

## Conclusion

In conclusion, our study highlights the sensitivity of BKMR to data distribution and covariance matrix assumptions, emphasizing the need for cautious interpretation of its results. Our findings highlight the importance of carefully considering data distribution when working with complex models like BKMR. While BKMR offers a valuable approach to understanding complex exposure-response relationships, its performance can be compromised in non-Gaussian data scenarios. Accounting for the effects of skewed data ensures more reliable, accurate, and consistent results. While the BKMR model provides a robust approach to understanding non-linear exposure-response relationships in diverse environmental health contexts, it is crucial to carefully evaluate its performance, especially regarding determining variable importance in non-Gaussian data scenarios.

The results of our study emphasize the possible difficulties associated with using a fixed PIP threshold and underscore the importance of using detailed criteria for variable importance determination, contributing to the ongoing effort to ensure the reliability and validity of understandings of the impacts of multi-pollutant mixtures on human health. Our future research will improve BKMR's robustness and explore alternative approaches to accommodate non-Gaussian data structures in environmental health studies. By addressing these challenges, we can enhance the reliability and validity of BKMR analysis and contribute to a deeper understanding of the impacts of multi-pollutant mixtures on human health.

**AUTHOR CONTRIBUTIONS**

All co-authors have meaningfully contributed to the production of this manuscript including funding, conception, data analysis, edition, revision, and final approval of the manuscript.

**CONFLICT OF INTEREST**

The authors declare no potential conflict of interests.